\documentclass{aa}
\usepackage{epsfig}

\newbox\grsign \setbox\grsign=\hbox{$>$} \newdimen\grdimen \grdimen=\ht\grsign
\newbox\simlessbox \newbox\simgreatbox \newbox\simpropbox \newbox\wtildebox 
\setbox\simgreatbox=\hbox{\raise.5ex\hbox{$>$}\llap
     {\lower.5ex\hbox{$\sim$}}}\ht1=\grdimen\dp1=0pt
\setbox\simlessbox=\hbox{\raise.5ex\hbox{$<$}\llap
     {\lower.5ex\hbox{$\sim$}}}\ht2=\grdimen\dp2=0pt

\newcommand{\greatsim}{\hbox{$\buildrel>\over\sim$}}
\newcommand{\lesssim}{\hbox{$\buildrel<\over\sim$}}
\newcommand{\Msun}{\mbox{$M_{\odot}$}}

\newcommand{\be}{\mbox{\begin{equation}}}
\newcommand{\ee}{\mbox{\end{equation}}}

\newcommand{\Mmin}{\mbox{$M_{\rm min}$}}
\newcommand{\Mmax}{\mbox{$M_{\rm max}$}}

\newcommand{\Mi}{\mbox{$M_i$}}
\newcommand{\Mp}{\mbox{$M$}}
\newcommand{\Mpl}{\mbox{$M_{l}$}}

\newcommand{\tdis}{\mbox{$t_{\rm dis}$}}

\newcommand{\Cref}{\mbox{$m_{\rm ref}$}}

\newcommand{\qev}{\mbox{$q_{\rm ev}$}}
\newcommand{\qevt}{\mbox{$q_{\rm ev}(t)$}}
\newcommand{\muev}{\mbox{$\mu_{\rm ev}$}}
\newcommand{\fsegr}{\mbox{$f_{\rm segr}$}}
\newcommand{\tsegr}{\mbox{$t_{\rm segr}$}}

\newcommand{\tgal}{\mbox{$t_0$}}

\newcommand{\muevt}{\mbox{$\mu_{\rm ev}(t)$}}

\newcommand{\nbody}{\mbox{$N$-body}}
\newcommand{\Nbody}{\mbox{$N$-body}}

\newcommand{\tdisbm}{\mbox{$t_{\rm dis}^{\rm BM}$}}

\begin{document}

\title{The photometric evolution of dissolving star clusters \\
 I: First predictions
}

\author{Henny J.G.L.M. Lamers \inst{1,2}, 
        Peter Anders \inst{3} and  Richard de Grijs\inst{4}
        }

\institute { 
                 {Astronomical Institute, Utrecht University, 
                 Princetonplein 5, NL-3584CC Utrecht, the Netherlands
                 {\tt  lamers@astro.uu.nl}}
            \and  {SRON Laboratory for Space Research, Sorbonnelaan 2,
                 NL-3584CC, Utrecht, the Netherlands}
            \and {Institut f\"ur Astrophysik, Georg-August-Universit\"at,
                  Friedrich-Hund-Platz 1, 37077 G\"ottingen, Germany
                  {\tt panders@astro.physik.uni-goettingen.de}}
            \and {Department of Physics \& Astronomy, University of Sheffield,
                  Hicks Building, Hounsfield Road, Sheffield S3 7RH, UK
                 {\tt R.deGrijs@sheffield.ac.uk}}
            }

\date{Received date ; accepted date}

\offprints{H. J. G. L. M. Lamers}

\abstract{ 
  We calculate the broad-band photometric evolution of unresolved star
  clusters, including the preferential loss of low-mass
  stars due to mass segregation, in a simplified way. 
The stellar mass function of a cluster
  evolves due to three effects: (a) the evolution of the massive stars
  reduces their number; (b) tidal effects before
  cluster-wide mass segregation reduce the mass function
  homogeneously, i.e. independently of the stellar mass; (c) after
  mass segregation has completed, tidal effects preferentially 
  remove the lowest-mass
  stars from the cluster. These effects result in a narrowing of the
  stellar mass range. These three effects are described
  quantitatively, following the results of $N$-body simulations, and
  taken into account in the calculation of the photometric history,
  based on the {\sc galev} cluster evolution models for solar metallicity
  and a Salpeter mass function. We find the following results:\\
  (1) During the first $\sim$ 40\% of the lifetime of a cluster its
  colour evolution is adequately described by the standard {\sc galev} models
  (without mass segregation) but the cluster gets fainter due to the
  loss of stars by tidal effects. During this phase the colour evolution is
  the same for clusters with and without initial mass segregation.\\
  (2) Between $\sim$ 40 and $\sim$ 80\% of its lifetime (independent of the
  total lifetime) the cluster gets bluer due to the loss of low-mass
  stars. This will result in an underestimate of the age of clusters
  if standard cluster evolution models are used. The correction
  increases from 0.15 dex for a cluster with a total lifetime of 3 Gyr
  to 0.5 dex for clusters with a total lifetime of 30 Gyr.\\
  (3) After $\sim$ 80\% of the total lifetime of a cluster it will rapidly
  get redder. This is because stars at the low-mass end of the main
  sequence, which are preferentially lost, are bluer than the AGB
  stars that dominate the light at long wavelengths. This will result
  in an overestimate of the age of clusters if standard cluster
  evolution models are used.\\
  (4) Clusters with mass segregation and the preferential loss of  low-mass stars 
  evolve along almost the same
  tracks in colour-colour diagrams as clusters without mass
  segregation. Therefore it will be difficult to distinguish this
  effect from that due to the cluster age for unresolved clusters.
  Only if the
  total lifetime of clusters can be estimated then the 
  colours can be used to give reliable age estimates.\\
  (5) The changes in the colour evolution of unresolved clusters
  due to the preferential loss of low-mass stars
  will affect the determination of the star
  formation histories of galaxies if they are derived from clusters
  that have a total lifetime of less than about 30 Gyr.\\
  (6) The preferential loss of low-mass stars 
  might explain the presence of old ($\sim$13 Gyr) clusters in
  NGC 4365 which are photometrically disguised as intermediate-age
  clusters (2 -- 5 Gyr), if the expected total lifetime of these clusters is
  between 16 and 32 Gyr. It may also explain the concentration of
  these clusters towards the center of NGC 4365.

}

\authorrunning{H.J.G.L.M. Lamers et al.}
\titlerunning{The photometric evolution of dissolving star clusters} 

\maketitle


\section{Introduction}
\label{sec:1}

The preferential loss of low-mass stars due to mass
  segregation in star clusters is not taken into
account in the cluster photometry models available in the literature.
However, this effect is of importance for the photometric evolution of
(``simple'') stellar population models (e.g. Leitherer et al. 1999; 
Bruzual \& Charlot, 1993; Anders \& Fritze-v. Alvensleben 2003)
 and hence the luminosity and
mass functions of cluster populations that are derived from the
analysis of observed spectral energy distributions (SEDs) 
of unresolved clusters compared to
standard models. In this paper we describe the expected effect of this
preferential loss of low-mass stars 
during the evolution of dissolving clusters, using a
simplified description for the dynamical mass-loss mechanism. We
compare and confirm this approach based on the results of {\it N}-body
simulations.

{\it N}-body simulations of clusters in tidal fields show that the
massive stars tend to concentrate towards the cluster center and the
low-mass stars preferentially populate the outer regions. This ``mass
segregation'' is commonly observed in open and globular clusters in
the Milky Way and in the Magellanic Clouds (e.g., de Grijs et
al. 2002a,b,c and references therein). 

An immediate and important consequence of mass segregation is that
low-mass stars tend to be more easily, and predominantly, ejected from
star clusters than high-mass stars (but see Brandl et al. 2001; de
Grijs et al. 2002a). Recently, Baumgardt \& Makino (2003; hereafter
BM2003) calculated the changes in the MF of clusters in the
tidal field of the Galaxy. They showed that the amplitude and shape of
the MF initially decrease almost homologously: the mass
distribution of the stars that survive stellar evolution decreases
with almost constant shape up to some critical time, after which the
mass distribution of the low-mass stars becomes very steep due to the
preferential loss of the lowest-mass stars. Their \nbody\
simulations of clusters at different galactocentric distances in the
Galaxy and with different orbits (circular vs. eccentric) show that
mass segregation becomes important when a cluster has lost a certain
fraction of its mass (approximately 70\%), almost independent of the
dissolution time-scale. In these simulations the role of binaries is
not taken into account, although binaries are important for mass
segregation (e.g., Inagaki \& Saslaw 1985; 
Bonnell \& Davies 1998; Portegies Zwart et al. 1999).
Therefore, in real clusters mass segregation
might occur earlier than predicted by the \nbody\ simulations of
BM2003. In some clusters mass segregation might even be established right
from the beginning (e.g. Hillenbrand 1997). 
We will refer to this as ``primordial mass segregation''.

In this paper we describe the expected effect of the preferential loss
of  low-mass stars, due to mass segeration, on
the integrated photometric evolution of clusters.
We use a simplified
method that enables us to explain and predict the effects for clusters
of widely different lifetimes (e.g. as derived by 
Lamers, Gieles \& Portegies Zwart 2005a).
 
The structure of this paper is as follows. In Sect. 2 we provide an
overview of the observational evidence for primordial and dynamical
mass segregation. We discuss the assumptions of our cluster models in
Sect. 3, i.e., we give a description of the decreasing mass of a
cluster due to the combination of stellar evolution and evaporation or
dissolution. 
In Sect. 4 we describe the way in which we treat the loss
of stars from the cluster due to stellar evolution and dissolution. 
In Sect. 5 we predict the photometric evolution of clusters,
taking the preferential loss of low-mass stars into account.
We compare the results with
those of the standard {\sc galev} simple stellar population models 
(Schulz et al. 2002;, Anders \& Fritze-v.-Alvernsleben 2003)
and describe the expected errors in
the age determinations of unresolved clusters. 
Sect. 6 contains a discussion and a suggestion of the a possible 
explanation for the
presence of an apparent intermediate age cluster population in the 
giant elliptical galaxy NGC 4365. The conclusions are in Sect. 7. 
 

\section{Observational evidence for mass segregation}
\label{sec:1a}

In the standard picture, stars in dense clusters evolve rapidly
towards a state of energy equipartition through stellar encounters,
with the corresponding mass segregation. The time-scale on which
dynamical mass segregation on cluster-wide scales is believed to occur
is the half-mass relaxation time-scale (e.g., Inagaki \& Saslaw
1985). However, observations of various degrees of mass segregation in
very young Galactic star clusters (e.g., Hillenbrand 1997; Testi et al. 1997;
Fischer et al. 1998; Hillenbrand \& Hartmann 1998; Hillenbrand \&
Carpenter 2000) suggest that at least some of this effect is related
to the process of star and star cluster formation itself (i.e. primordial
mass segregation):
these clusters are often significantly younger than
their two-body relaxation time (even the equivalent relaxation time in
the core). 

The same effect is found in several Magellanic Cloud clusters.
Ground based studies of several rich compact LMC clusters, e.g. NGC
2100 (Westerlund 1961); 
NGC 2098 and SL 666 (Kontizas et
al. 1998; Gouliermis et al. 2004), show strong 
indications of mass segregation. In addition, 
observations with the {\sl HST} have also
resulted in convincing cases for mass segregation in Magellanic Cloud
star clusters: e.g., Fischer et al. (1998); Elson et al. (1999); Santiago
et al. (2001); de Grijs et al. (2002a,b,c) and Sirianni et al. (2002).

The clearest evidence for primordial mass segregation 
is found in the Orion Nebula Cluster (ONC), NGC 3603 and R136.
Hillenbrand (1997) and Hillenbrand \& Hartmann (1998) presented 
clear evidence for mass segregation in the ONC
for the $m > 5 M_\odot$ component, with some evidence for general mass
segregation down to $m \simeq 1$--$2 M_\odot$ (see the review by
Larson 1993). 
 Ground and {\sl
HST}-based observations of NGC 3603, one of the few moderately  massive young star clusters
in the Galaxy, have
shown strong evidence for mass segregation 
in this very young ($1\pm1$ Myr-old) cluster  (N\"urnberger \&
Petr-Gotzens 2002; Stolte et al. 2004; Sung \& Bessell 2004; 
Grebel 2004).
The cluster R136 in the 30 Doradus star-forming region in the LMC
(age $\lesssim$ 3--4 Myr, cf. Hunter et al.  1995), has been studied
extensively, both from the ground and with the {\sl HST}. A variety of
techniques have revealed strong mass segregation, certainly for radii
$r \lesssim 0.5$ pc (e.g., Campbell et al. 1992; Larson 1993; Malumuth
\& Heap 1994; Hunter et al. 1995; Brandl et al. 1996).

These results suggest that some initial mass segregation may occur
already during the formation of the cluster.

\subsection{Disentangling dynamical from primordial effects}

To disentangle the effects of dynamical versus primordial mass
segregation, de Grijs et al. (2002c) studied six  rich LMC clusters 
of different ages and masses.
All clusters in their sample show clear evidence
of mass segregation: (i) the luminosity function slopes steepen with
increasing cluster radius, and (ii) the brighter stars are
characterized by smaller core radii. More importantly, for all sample
clusters, both the slope of the luminosity function in the cluster
centres and the degree of mass segregation were found to be similar to
each other, within observational errors of a few tenths of power-law
slope fits to the data. They conclude that the {\it
initial} MFs must have been very similar, down to $\sim 0.8
- 1.0 M_\odot$.

Upon closer inspection of these data, however, we do 
notice significant differences
among the degrees of mass segregation among the clusters, which may
have been caused by dynamical mass segregation effects (for the older
objects), in addition to probable primordial mass segregation. In
Fig. 5 of de Grijs et al. (2002c) the behaviour of the slope of the 
stellar luminosity function as a function of
radius (which steepens outward), 
for different age
bins, are found to scale as $1 : (0.8 \pm 0.2) : (0.2 \pm 0.1)$ for
ages of $10^7 : 10^8 : 10^9$ yr. This implies that the 
mass segregation due to dynamical effects becomes significant 
on time-scales of
a few $\times 10^8$ yr. Dynamical mass segregation is expected to
occur on the half mass relaxation time. Our 
empirical estimate of the time at which dynamical mass segregation occurs
is consistent with, or perhaps
relatively short compared to, estimates for the half-mass relaxation
time-scales of Galactic GCs (e.g., Gnedin \& Ostriker 1997) which
might be caused by differences in the relaxation times.

Combining the results of the studies of very young  clusters 
 with 
the observed changes in the radial dependence of the slope of the mass
function with age,
we may conclude that (most, if not all) clusters are formed with primordial
mass segregation and that the mass segregation increases on 
a half-mass relaxation time of a few $10^8$ yr.
This agrees with the results of a recent study of mass segregation
in very young open clusters in the solar neighbourhood by Schilbach  et
 al. (2006). In the group of the youngest clusters, with ages between
5 and 30 Myr, They found about as many clusters with  
as without initial mass segregation. 


\section{Mass loss from a cluster by stellar evolution and dissolution}

\begin{figure}
\centerline{\psfig{figure=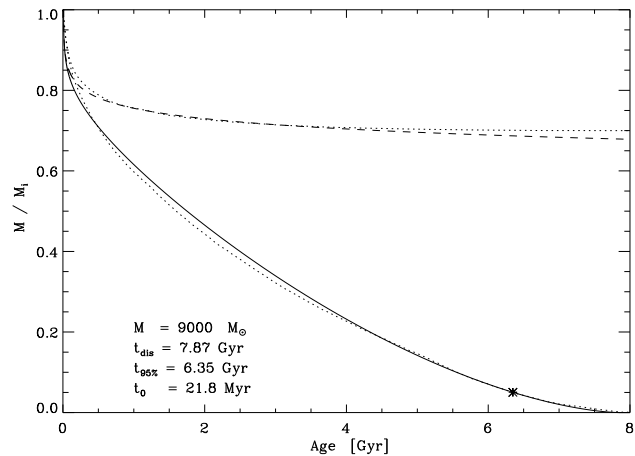,width=9.0cm}}
\caption[]{The decreasing mass of a cluster with an inital mass of
  9000 \Msun, predicted by Eq. (\ref{eq:muapprox}) with $\tgal=21.8$
  Myr. The solid line is the decreasing mass due to stellar evolution
  and dissolution, the dashed line shows the decreasing mass if stellar
  evolution were the only mass-loss mechanism.
  This agrees very well  with the \nbody\ simulations by BM2003
  of a cluster of 9000 \Msun\ in a circular orbit at a galactocentric
  distance of 8.5 kpc with an initial concentration parameter of
  $W_0=5$ (dotted lines). The asterisk indicates the moment when 95\% of the
  initial mass is lost. This is the value of \tdisbm.
}
\label{fig:9000msun}
\end{figure}

In order to describe the expected effects of changes in the mass
function of clusters due to mass segregation and the preferential
loss of low-mass stars, we first describe the overall mass loss of a
cluster by stellar evolution and dissolution.

The loss of mass due to stellar evolution and to dissolution can be
described as

\begin{equation}
\frac{{\rm d}\Mp}{{\rm d}t}=  
   \left(\frac{{\rm d}\Mp}{{\rm d}t}\right)_{\rm ev} +
   \left(\frac{{\rm d}\Mp}{{\rm d}t}\right)_{\rm dis} \quad ,
\label{eq:dmdt}
\end{equation}
The mass loss due to stellar evolution is 

\begin{equation}
\left(\frac{\rm d \Mp}{{\rm d}t}\right)_{\rm ev} = 
\Mi \frac{{\rm d} \muev(t)}{{\rm d}t}=
\frac{\Mp}{\muev(t)} \frac{{\rm d} \muev(t)}{{\rm d}t} \quad ,
\label{eq:dmdtev}
\end{equation}
where $1-\muevt=\qevt$
is defined as the mass fraction of the initial mass that is lost due to
stellar evolution. Thus, $\Mp (t)=\Mi \muevt$ if there is no dissolution.
The function $\qevt$ can be expressed as the
simple approximation  

\begin{equation}
\log \qev(t)= (\log t-a_{\rm ev})^{b_{\rm ev}}+c_{\rm ev}~~{\rm for}~~
t>12.5~{\rm Myr}
\label{eq:qevgot}
\end{equation}
The values of $a_{\rm ev}$, $b_{\rm ev}$ and $c_{\rm ev}$ 
were derived by Lamers et al. (2005b) from the {\sc galev} cluster
evolution models for different metallicities, $0.0004 \le Z \le 0.05$.
 For solar metallicity they found $a_{\rm ev}=7.00$,
$b_{\rm ev}=0.255$ and $c_{\rm ev}=-1.805$.  Equation (\ref{eq:qevgot})
describes the mass-loss fraction of the models at $t>12.5$ Myr, with
an accuracy of a few per cent. The mass loss at younger ages is
negligible because the most massive stars with $M_*>30 \Msun$ hardly
contribute to the mass of the cluster.

The mass loss due to dissolution can be written as

\begin{equation}
\left(\frac{{\rm d}\Mp}{{\rm d}t}\right)_{\rm dis} = -
\frac{\Mp}{\tau_{\rm dis}} = 
               - \frac{\Mp}{\tgal \Mp^{\gamma}}= 
               - \frac{\Mp^{1-\gamma}}{\tgal} \quad ,
\label{eq:dmdtdis}
\end{equation}
where the instantaneous dissolution time-scale $\tau_{\rm dis}$ 
is assumed to depend as a power law 
on the mass of the cluster, as $\tau_{\rm dis}= \tgal M^{\gamma}$.
Boutloukos \& Lamers (2003)
and Lamers et al. (2005a) derived the value $\gamma$ empirically
from a statistical study of cluster samples in different galaxies
and found it to be $0.62 \pm 0.05$.
BM2003 derived the same value of $\gamma=0.62$ from their
{\it N}-body simulations of clusters in tidal fields,
 in excellent agreement with the empirical studies.
The constant $\tgal$ depends on the tidal field at the
location of the cluster.

Lamers et al. (2005b) have shown that the numerical solution of
Eq. (\ref{eq:dmdt}) can be approximated to a high degree of accuracy
by the analytical expression

\begin{equation}
 M(t)\simeq M_i \times
\left\{(\muevt)^{\gamma}-\frac{\gamma t}{\tgal \Mi^{\gamma}}
\right\}^{1/\gamma} \quad ,
\label{eq:muapprox}
\end{equation}
where $\Mi$ is the initial mass of the cluster.

Equation (\ref{eq:muapprox}) is valid if the first term on
the right-hand side is greater than the second term. In the opposite
situation, i.e., when the mass lost through dissolution is larger than
the mass remaining after mass loss via stellar evolution, then
$M(t)=0$ and the cluster is completely dissolved. Lamers et
al. (2005b) compared this expression for the mass evolution of star
clusters with the results of the \nbody\ simulations of BM2003, and found
that the agreement is extremely good, i.e., within a few per cent over
more than 98\% of the lifetime of the clusters.

The total dissolution time of clusters can be derived from
Eq. (\ref{eq:muapprox}). Lamers et al. (2005b) showed that it is
approximately

 \begin{eqnarray}
\tdis(\Mi) &=& 6.60 \times 10^2 \left(\frac{\Mi}{10^4}\right)^{0.653} \times
\nonumber \\
 & & \tgal^{0.967-0.00825\times \log(M_i/10^4)}
\label{eq:ttot}
\end{eqnarray}
for solar metallicity and $\gamma=0.62$. They also showed that the
time when 95\% of the initial mass is lost,
which is the dissolution time-scale adopted by BM2003, \tdisbm,
is approximately 0.8\tdis.

Fig. \ref{fig:9000msun} shows the decreasing mass of a cluster with an
initial mass of 9000 \Msun, described by Eq. (\ref{eq:muapprox}).
The value of $\tgal=21.8$ Myr was chosen in such a way that the
cluster has lost 95\% of its mass in 6.35 Gyr, which is the value
derived by BM2003 from \nbody\ simulations for a cluster of 9000 \Msun\ with
an initial concentration parameter of $W_0=5$ 
in a circular orbit at $R_{\rm gal}=8.5$ kpc. The total dissolution
time of such a cluster is $\tdis=7.87$ Gyr, which is 24\% greater than \tdisbm.
Fig. \ref{fig:9000msun} shows that Eq. (\ref{eq:muapprox}) agrees very
well with the result from $N$-body simulations.


\section{The changing mass function of a cluster}

In the previous section we described the decreasing mass of the
cluster due to evolutionary and dynamical processes. In this section,
we describe how stellar mass loss and cluster dissolution change the
{\it stellar MF} of the cluster during its evolution.


\subsection{The changes in the mass function predicted by \Nbody\
  simulations}

BM2003
calculated the dynamical evolution of a large grid of
clusters in the tidal field of the Galaxy by means of $\nbody$
simulations.
The clusters have different masses and
different initial density distributions which are expressed in terms
of King profiles with central concentration parameters $W_0=5.0$ or
7.0. The clusters orbit at different galactocentric distances, from 
2.8 to 15 kpc, in circular or elliptical orbits. 
The initial masses of these cluster models are in the range 
from 4000 to 70000 \Msun.
An initial Kroupa (2001) MF of the type
$N(M){\rm d}M=M^{-\alpha}{\rm d}M$ with $\alpha=2.3$ for $0.5 \le M \le 100 \Msun$ and 
$\alpha=1.3$ for $0.1 \le M< 0.5 \Msun$ was adopted.
BM2003 defined the dissolution time of their models as the age at which
only 5\% of the initial mass remains due to stellar evolution and
dissolution.  We will refer to these dissolution times as \tdisbm.
The dissolution times of these models
are in the range $2.3 < \tdisbm < 40$ Gyr. 

The \nbody\ simulations of BM2003 show several remarkable features:\\
(a) The preferential loss of low-mass stars starts at about the same
fractional age of the cluster, i.e. at $\tsegr \simeq 0.20~\tdis$.\\
(b) Before this age the fractional decrease of the number of stars  
is almost independent of mass (except for the most massive stars whose
number decreases because of stellar evolution).  This is because stars
of almost all masses can be kicked out by encounters with the most
massive ones. \\
(c) After \tsegr\ the cluster mainly loses high-mass stars, due to
stellar evolution, and low-mass stars due to evaporation, but almost no stars
of intermediate age. \\
(d) The changes in the mass function at the low mass end, $M \le
2\Msun$,  as a function of $t/\tdis$ are very similar for all models,
despite large differences in the initial cluster mass, dissolution time
and ellipticity of the orbit of the cluster. 


\subsection{Simplified models: the concept}

\begin{figure}
\centerline{\psfig{figure=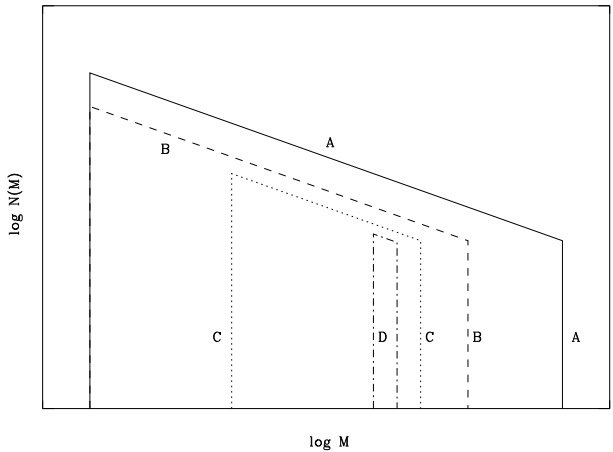,width=9.0cm}}
\caption[]{Schematic changes in the stellar MF, $\log(N)$
  versus $\log(M)$, of a cluster due to stellar evolution and
  dissolution. The initial MF is assumed to be a power law.
  A: the inital MF. B: Stellar evolution
  has removed the most massive stars and dynamical effects, before
  cluster-wide mass segregation, have reduced the number of stars 
  of all masses with equal probability. C: after mass
  segregation has occurred dynamical effects have removed the lowest-mass
  stars. D: the MF just before the cluster is
  completely dissolved.}
\label{fig:concept}
\end{figure}

Based on the results of the \nbody\ simulations, we can derive a
simple model that agrees with the basic results of BM2003 and that
allows us to subsequently calculate the photometric evolution of dissolving 
clusters.

We describe the effects of stellar evolution and
cluster dissolution on the stellar MF 
by the following approximations:\\
(1) Stellar evolution removes stars at the high-mass end of the 
MF but leaves the rest of the mass distribution unchanged.
The stellar upper mass limit decreases with time. \\ 
(2) Initially, dissolution will remove stars
of all masses with about equal probability. This means that the
number of stars of all masses will decrease but the slope of the mass
distribution will remain unchanged. This agrees with the results of
the \nbody\ simulations.\\ 
(3) When mass segregation has occurred,
dissolution preferentially removes low-mass stars from the
periphery of the cluster.
 In our simple model we will assume that in this phase,  i.e. at $t>t_{\rm segr}$,  
the dynamical effects remove only  the stars with the 
lowest remaining mass from the cluster.
Thus,  the lower mass limit 
of the cluster stars will increase to higher values.
Stellar evolution will continue to reduce the upper mass limit during
all phases. (In fact, the evolution of stars of all masses is fully
taken into account in our method because we use the results from 
the {\sc galev} cluster evolution models.)\\
(4) The transition between phases (2) and (3) is likely to be gradual.
Therefore, we assume that a fraction $1-\fsegr$ of the mass that is lost due to
dynamical effects is independent of the mass and the 
fraction $\fsegr$ is lost in the form of low-mass stars only. The factor
\fsegr\ can be specified as a function of time and increases from some initial value
$f_0 \equiv \fsegr (t=0)$ to $\fsegr=1$ at $t>\tsegr$. 
If there is no primordial mass segregation then $f_0=0$.

These assumptions have the advantage that {\it the slope of the mass
distribution remains constant} during all phases of a cluster's
evolution. Only the maximum and minimum stellar mass (due to effects 1
and 3, respectively), and the constant describing the total number of
stars (due to effect 2) change as a function of time. With these
simplifications, we can study and understand the expected
changes in the photometry of star clusters during their
evolution. Figure \ref{fig:concept} shows the concept of our
approximations. In this simple description the MF gets
narrower with time and reaches a single mass just before the cluster
is fully dissolved.

Our assumptions are in  agreement with those of the \nbody\
simulations of the dynamical evolution of clusters in tidal fields
by BM2003 (see Sect. 4.1). The main
difference is in the shape of the mass function at the low-mass end.
BM2003 showed that the MF at $M<2 \Msun$ becomes steeper with time,
i.e. ${\rm d} \log(N)/{\rm d} \log(M)$ becomes more positive after core collapse.
In our model we approximate this by a MF that keeps its original slope
but with a lower mass limit that shifts to higher $M$ as time progresses.
The total amount of mass at $M < 2 \Msun$ is the same in both the BM2003
models and in our models.


\subsection{Changes in the mass function due to stellar evolution}
\label{sec:3.2}

We adopt a power-law stellar IMF, $N(M){\rm d}M = C(t) M^{-\alpha}{\rm
  d}M$
with $\alpha=2.35$, i.e., a Salpeter IMF, in the range of $\Mmin(t) <
M < \Mmax(t)$, with $\Mmin(0)=0.15$ and $\Mmax(0)=85$ \Msun. With this value
of $\Mmin(0)$ the mean stellar mass in a cluster is very similar to that
of the Kroupa (2001) MF. 
The factor $C(t)$ is related to
the luminous mass of the cluster at time $t$ via

\begin{equation}
\Mpl(t)=\frac{C(t)}{\alpha-2}~
\{\Mmin(t)^{2-\alpha}-\Mmax(t)^{2-\alpha}\} \quad .
\label{eq:mmaxt}
\end{equation}

Stellar evolution reduces the mass of the cluster because of the
evolution of its most massive stars. We will assume that massive stars
keep their initial mass until they reach the end of their life. This
is a reasonable assumption because the dominant mass loss for our study
occurs after about 10 Myr, when stars with $M>12 \Msun$ no longer
exist. Stars with $M<12\Msun$ have no appreciable mass loss during the
main-sequence phase, and so their mass loss occurs in the relatively
short time of less than about 10\% of the main-sequence lifetime at
the end of their life. For $\Mmax(t)$ we adopt the main-sequence turn-off
lifetime of stars given by Hurley et al. (2000).


\subsection{Changes in the mass function due to dissolution}
\label{sec:3.3}

\begin{figure*}
\vspace{-4cm}
\centerline{\psfig{figure=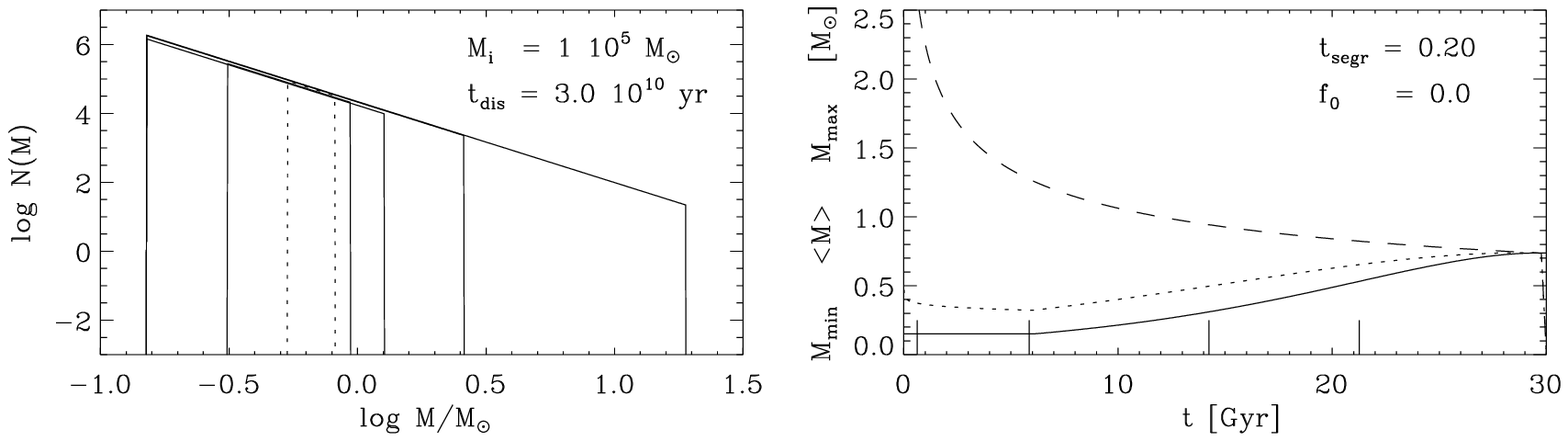,width=12.0cm}}
\vspace{-5.5cm}
\centerline{\psfig{figure=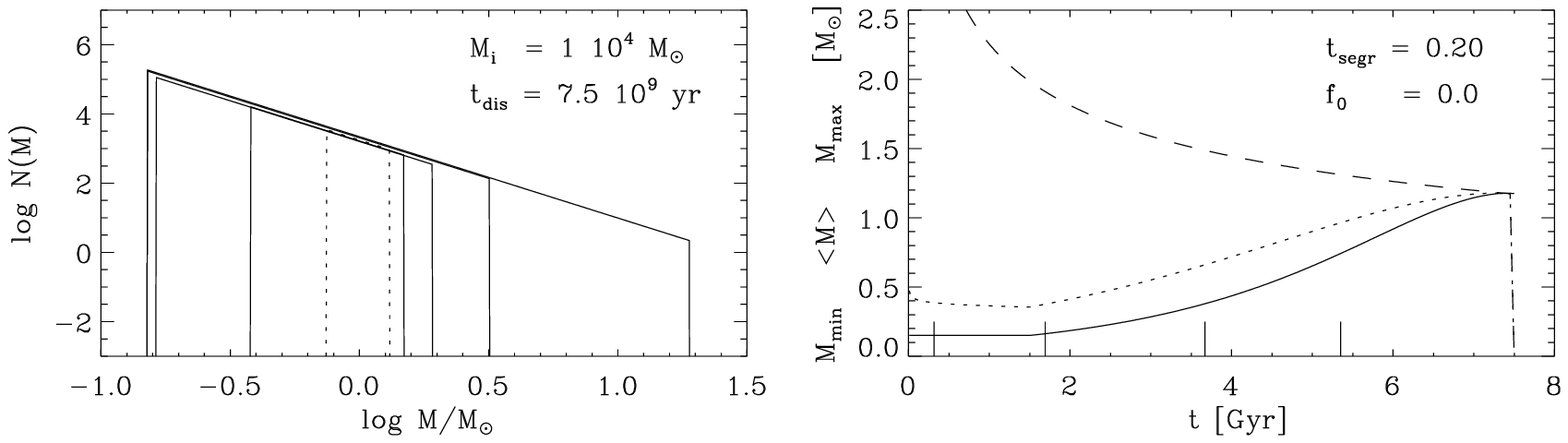,width=12.0cm}}
\vspace{-5.5cm}
\centerline{\psfig{figure=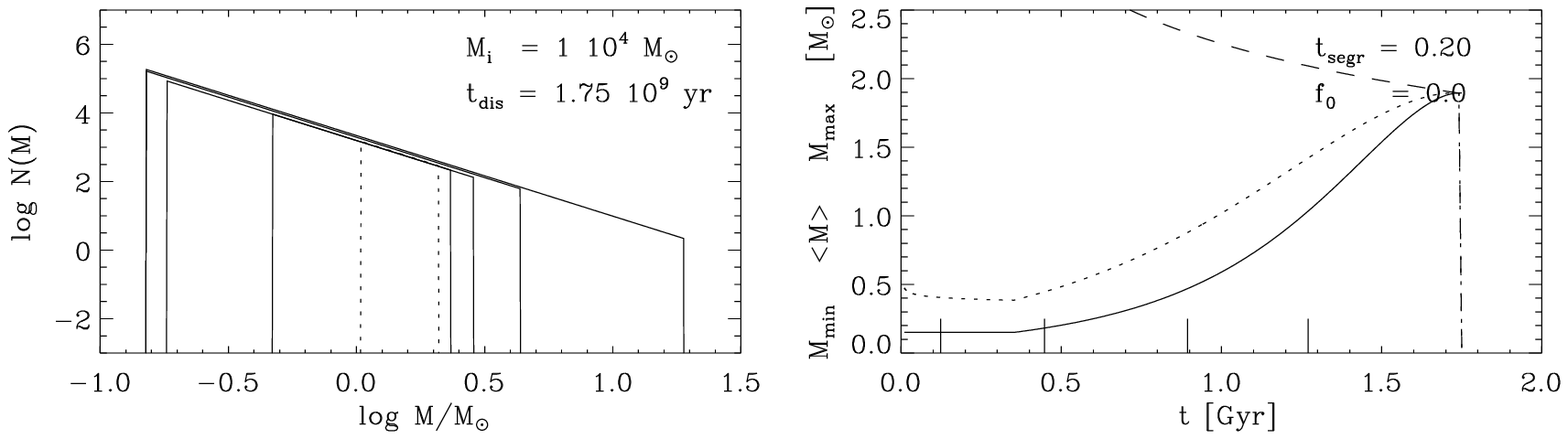,width=12.0cm}}
\caption[]{The changes in the stellar mass function (excluding
  remnants) of dissolving
  clusters with a total dissolution time of $\tdis=30$, 7.5 and
  1.75 Gyr (from top to bottom). The preferential loss of low-mass
  stars sets in at $\tsegr=0.20~\tdis$. The left-hand panels show the mass
  function when the mass of the cluster is 100, 75, 50, 25 and 10\% of
  its initial mass \Mi\ (from outer to inner envelope). The dotted line
  shows the MF when $M(t)=0.10~\Mi$. The right-hand panels
  show the changes in the maximum (dashed line), minimum (solid line)
  and mean (dotted line) stellar mass as a
  function of time for the same models. The vertical tick marks indicate the times when
  the mass of the cluster is reduced to 75, 50, 25 and 10\% of \Mi\, respectively
  (from left to right).
}
\label{fig:simple-models}
\end{figure*}
\begin{figure*}
\vspace{-4cm}
\centerline{\psfig{figure=mf_me5_tdis30_tseg02_fseg0.ps,width=12.0cm}}
\vspace{-5.5cm}
\centerline{\psfig{figure=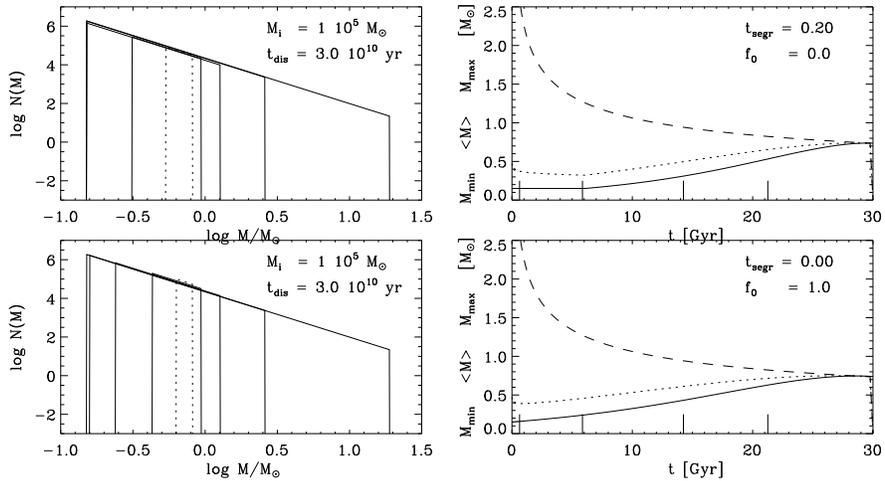,width=12.0cm}}
\caption[]{The changes in the mass function of a cluster with and
without initial mass segregation (top and bottom, respectively),
plotted in the same way as in Fig. 3.
Notice the difference in the mass function at the low-mass end.
}
\label{fig:simple-models-b}
\end{figure*}

If a fraction $1-\fsegr$ of the mass lost by dissolution is 
independent of the stellar mass and the fraction $\fsegr$ is lost 
in the form of the lowest-mass stars, then
we can find an
expression for the decrease of the stellar minimum mass with time by
combining Eqs. (\ref{eq:muapprox}) and (\ref{eq:mmaxt}). This yields

\begin{eqnarray}
\frac{{\rm d} M_{\rm min}^{2-\alpha}}{{\rm d}t} &=&
\frac{(\alpha-2)\fsegr}{C(t)} \left(\frac{{\rm d} \Mp}{{\rm d}t}\right)_{\rm dis}
\nonumber \\ & =& \frac{(\alpha-2)\fsegr}{C(t)~ \tgal} \Mp^{1-\gamma} .
\label{eq:dmmindt}
\end{eqnarray}
 Similarly, by combining
Eq. (\ref{eq:dmdtdis}) with the derivative of Eq. (\ref{eq:muapprox}) we
find an expression for the decrease of the constant $C(t)$ of the mass
function,

\begin{equation}
\frac{{\rm d} \ln C(t)}{{\rm d}t} = (1-\fsegr) ~ \left(\frac{{\rm d}
\ln \Mp}{{\rm d}t}\right)_{\rm dis} = \frac{1-\fsegr}{\tgal}
\Mp^{-\gamma} \quad .
\label{eq:dCdt}
\end{equation}
For any time-dependent segregation parameter, $\fsegr(t)$, we can
calculate the evolution of the MF of a cluster by solving
Eqs. (\ref{eq:mmaxt}), (\ref{eq:dmmindt}) and (\ref{eq:dCdt}).


\subsection{Simple models}

To demonstate the effect of mass segregation and the preferential loss
of low-mass stars on the time-dependent MF
we assume that \fsegr\ increases in time as a step function (in
agreement with the results of the \nbody\ simulations).

\begin{eqnarray}
\fsegr&=& 0 ~~~~~~~~~~~ {\rm if}~~ t<\tsegr \nonumber  \\
\fsegr&=& 1 ~~~~~~~~~~~ {\rm if}~~ t\ge\tsegr 
\label{eq:fsegr} 
\end{eqnarray}
where $\tsegr$ is the time when cluster-wide mass segregation has
occurred. 
The models of BM2003 suggest that $\tsegr \simeq 0.20~\tdis$ if there is
no intial mass segregation. However, observations suggest that
clusters may be born with initial mass segregation, in which case
$\tsegr=0$. We consider models with $\tsegr=0$ and $\tsegr = 0.2~\tdis$. 

Figure \ref{fig:simple-models} shows the changes in the stellar mass function
(without the remnants)
of clusters with different dissolution times between 1.75 and 30 Gyr. 
The longer the dissolution
time, the smaller the mass of the most massive stars at the end of the cluster's
life. The lower-mass limit starts shifting to higher masses only after
$\tsegr$, which is assumed to be $0.20~\tdis$ in these models. The
mean stellar mass at the end of the cluster's life is equal to the
mass at the main-sequence turn-off at $t=\tdis$. By that time, the
lower-mass limit has increased to that same value. The mean mass first
decreases from its initial value of 0.545 \Msun\ due to the rapid loss 
of the massive stars to about 0.35 \Msun , and increases after \tsegr\
due to the preferential loss of low-mass stars. The mean mass at the
end of the cluster's lifetime is 0.75, 1.2 and 2.9 \Msun\ for
dissolution times of 30, 7.5 and 1.75 Gyr, respectively. 

The mass functions are independent of the adopted initial cluster
mass, and only depend on \tdis\ and \tsegr.
The evolution of the mean mass for the model with $\tdis=30$ Gyr 
resembles that predicted by BM2003 for a cluster of $7.1 \times 10^4$
\Msun\
with $\tdisbm=23.7$ Gyr and $\tdis=30.2$ Gyr (BM2003, Fig 15). 
The difference is significant mainly near the very end of the 
cluster's life, after 24 Gyr,
where BM2003 predict a rapid increase in the mean mass to
1.35 \Msun\ compared to our value of 0.75 \Msun. This is because the
mass function calculated by BM2003 also includes the remnants of the
massive stars, mainly neutron stars of about 1.5 \Msun.
Our mass function is for the luminous stars only.
 
Figure \ref{fig:simple-models-b} shows the effect of initial
mass segregation on the evolution of the stellar mass function of
clusters.
The top panel is for $\tsegr=0.20~\tdis$ and the lower panel is for
$\tsegr=0$, i.e. for initial mass segregation. The difference results in 
a different evolution of the mass function at the
low-mass end. In case of initial mass segregation, the mean mass hardly
decreases during the early phases, because the rapid loss of high-mass
stars due to stellar evolution is more than compensated for by the
simultaneous loss of low-mass stars.

\begin{figure*}
\centerline{\psfig{figure=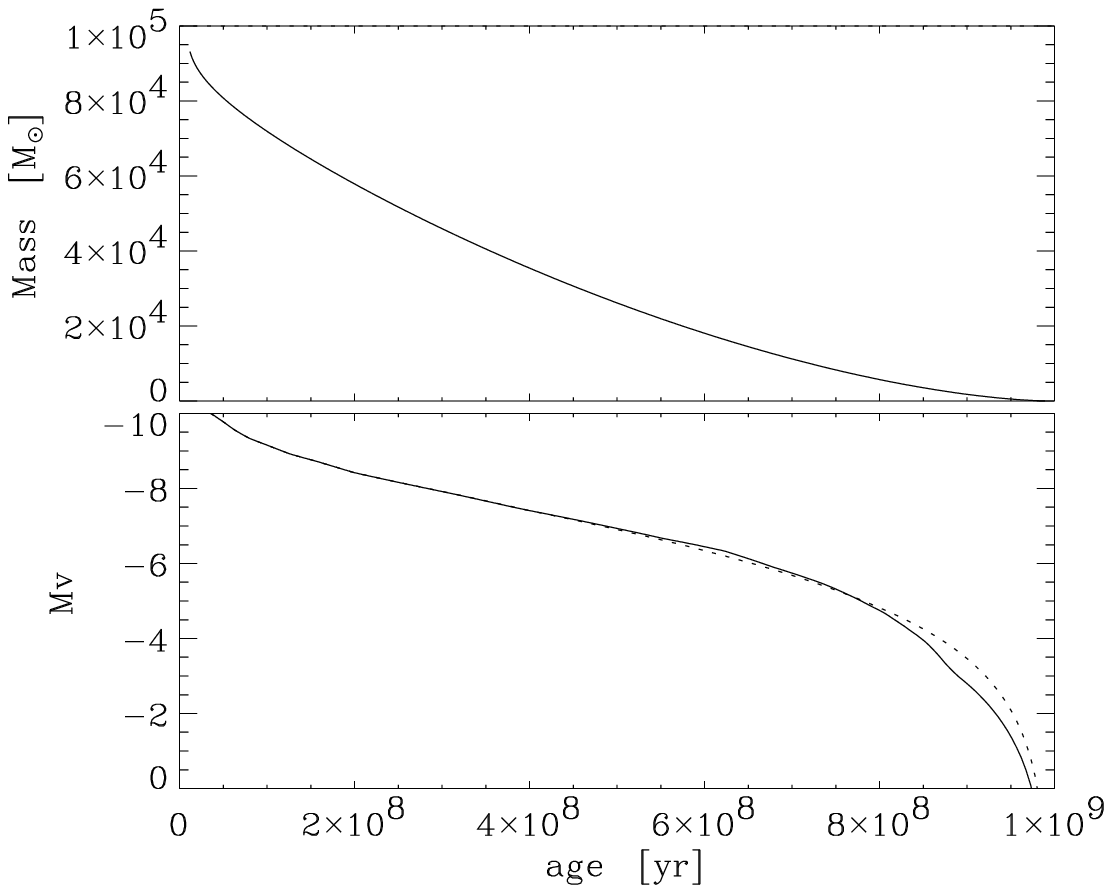,width=10.5cm}
            \hspace{-4.0cm}\psfig{figure=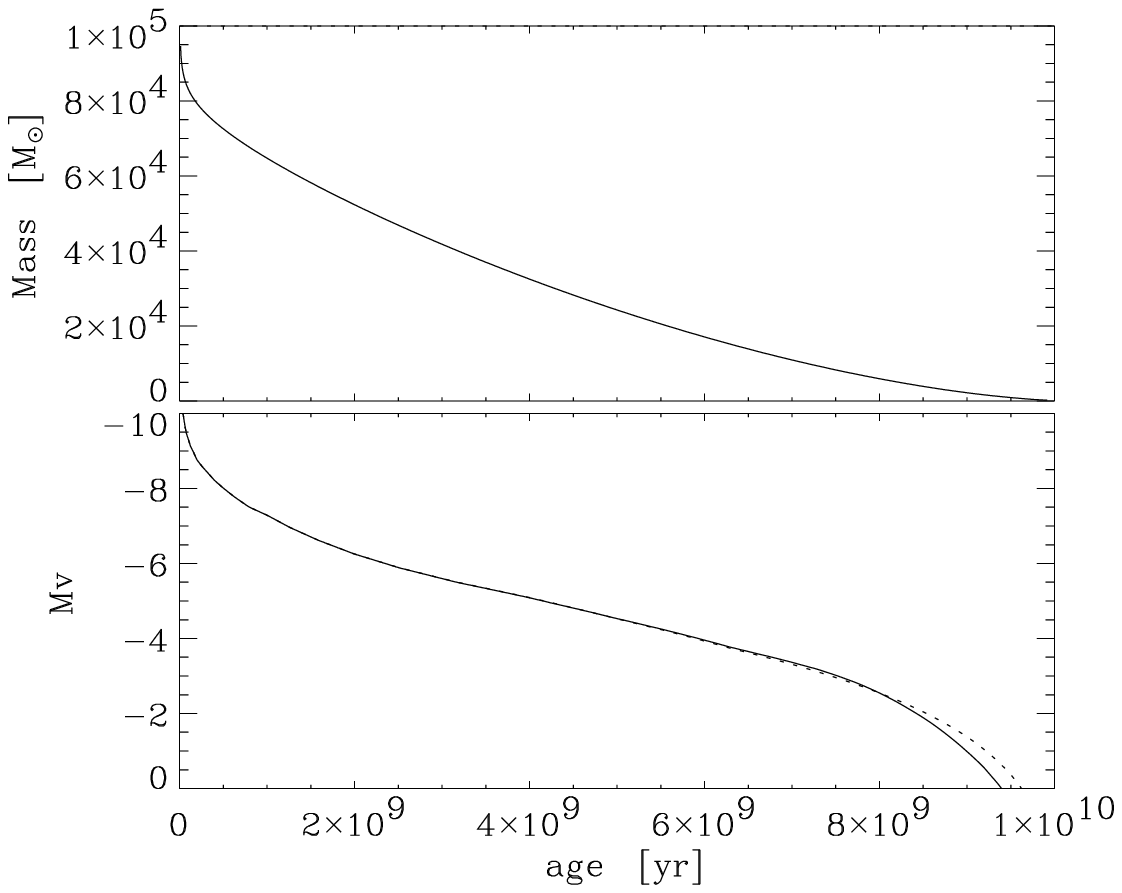,width=10.5cm}}
\vspace{-2.0cm}
\centerline{\psfig{figure=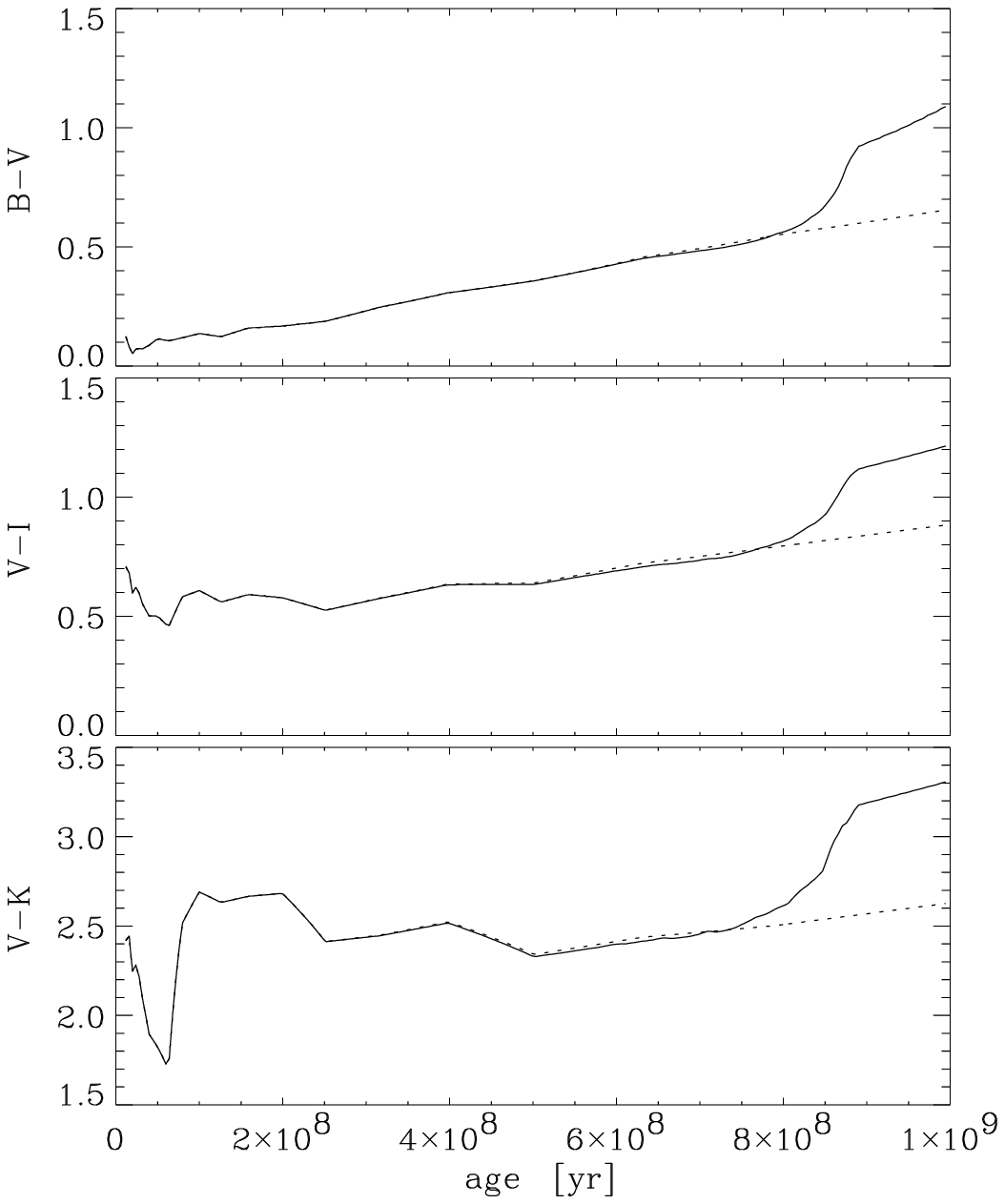,width=10.5cm}
            \hspace{-4.0cm}\psfig{figure=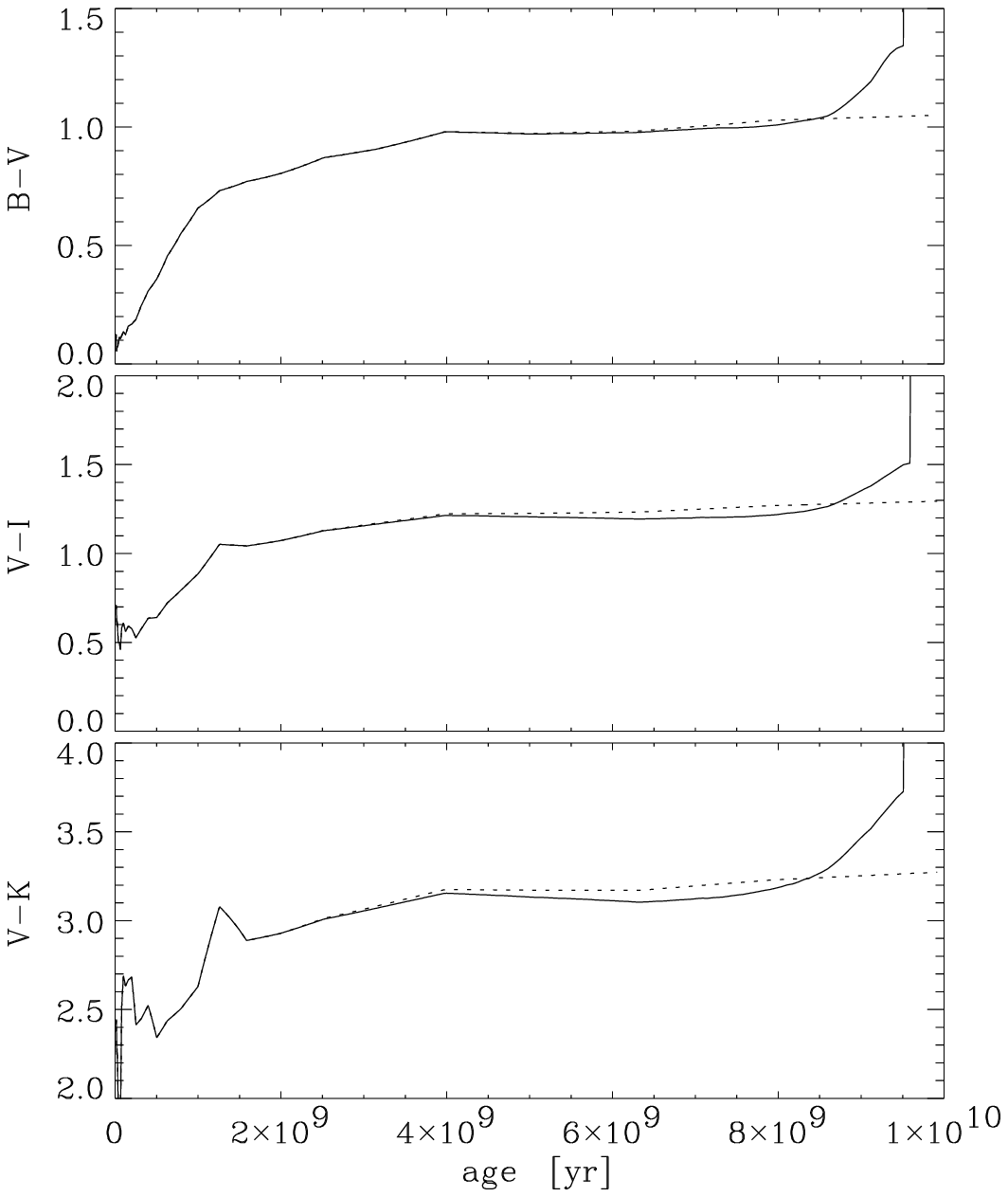,width=10.5cm}}
\caption[]{The mass, $M_V$, and colour history of dissolving
            star clusters with mass segregation and the preferential
            loss of low-mass stars taken into account. 
Left-hand figure: a cluster with an initial mass of $10^5~\Msun$, 
an initial Salpeter stellar IMF, no
initial mass segregation and a dissolution time of 1 Gyr. The right-hand
figure is for a similar model but with $\tdis$=10 Gyr.
The upper panels show the changes in mass and $M_V$ and the lower panels
the changes in {\sl HST}/WFPC2 broad-band colours. The solid lines show the 
photometric history of dissolving clusters with the preferential 
loss of low-mass stars, as described in Sect. 4. 
The dashed lines show the photometry
of dissolving clusters without the preferential loss of low-mass
stars (i.e. the standard model).}
\label{fig:model_mass_phot}
\end{figure*} 


\section{The photometric history of dissolving clusters}

In this section we calculate the photometric history of
dissolving star clusters based on our description of the changes in
the mass function.
We adopt the {\sc galev} models  with a Salpeter IMF
of exponent $-2.35$. The change in the upper mass limit due to 
stellar evolution is included in the {\sc galev} models.
The changes in the lower mass limit are described in Sect. 4.
The {\sc galev}  models are based on the stellar evolution
calculations of the Padova group and include, amongst others,  a description 
of the AGB evolution. We will see that these AGB stars are
important for the colours of old clusters.

We calculated models with and without initial mass segregation
($\tsegr=0.0$ and 0.2 
respectively) in Eq. (\ref{eq:fsegr})),
 and with total dissolution times of
0.1, 0.3, 1, 3, 10 and 30 Gyr. The results are shown in 
Fig. \ref{fig:model_mass_phot} for two models with $\tdis=1$ Gyr (left) 
and 10 Gyr (right).

The upper panels of Fig. \ref{fig:model_mass_phot} show the changes in the mass and visual
magnitude of a cluster with an initial mass of $10^5~\Msun$.
The lower panels show
the changes in the photometric history in $B-V$, $V-I$ and $V-K$. The
solid lines show the prediction including the
loss of low-mass stars but without initial segregation ($\fsegr=0$, 
$\tsegr=0.2$). The dotted lines show the results of the standard
{\sc galev} models, i.e. without the preferential loss of low-mass stars. 
We note that the colours are independent of the adopted
initial mass of the cluster, but $M_V$ has to be scaled to the
decreasing cluster mass.

Fig. \ref{fig:model_mass_phot} shows that near the end of the cluster's life the
visual luminosity becomes fainter than in case of no preferential loss
of low-mass stars. 
This is because of the missing contribution
from the large number of lost low-mass main-sequence stars.

The colour evolution of models with preferential loss of low-mass stars
is different from that of the standard models. At ages between about 
$0.4 < t/\tdis < 0.8$,
clusters that include mass segregation get slightly
{\it bluer} than clusters without mass segregation (lower panels of
Fig. 5). This effect is  
stronger for clusters with increasing values of \tdis. This means that
it will be stronger for more massive clusters, which have a longer
lifetime, than for low-mass clusters. It is due to the loss of red low-mass 
main-sequence stars.
The situation changes drastically after $t \simeq 0.8~\tdis$ when the
clusters with mass segregation  get much {\it redder} than those
without mass segregation. This is due to the fact that 
late in the lifetime of a cluster its red colour is dominated by AGB stars.
Stars at the low-mass end of the main sequence are bluer than 
AGB stars, so the loss of stars at the low-mass end of the main sequence 
makes the cluster redder.
{\it The blueing of the clusters at $0.4 \le t/\tdis \le 0.8$ implies
that the age of clusters derived from standard cluster evolution
models will be underestimated, whereas the age will
be overestimated for clusters with $t \ge 0.8~\tdis$.}

\begin{figure}
\centerline{\hspace{+4.0cm}\psfig{figure=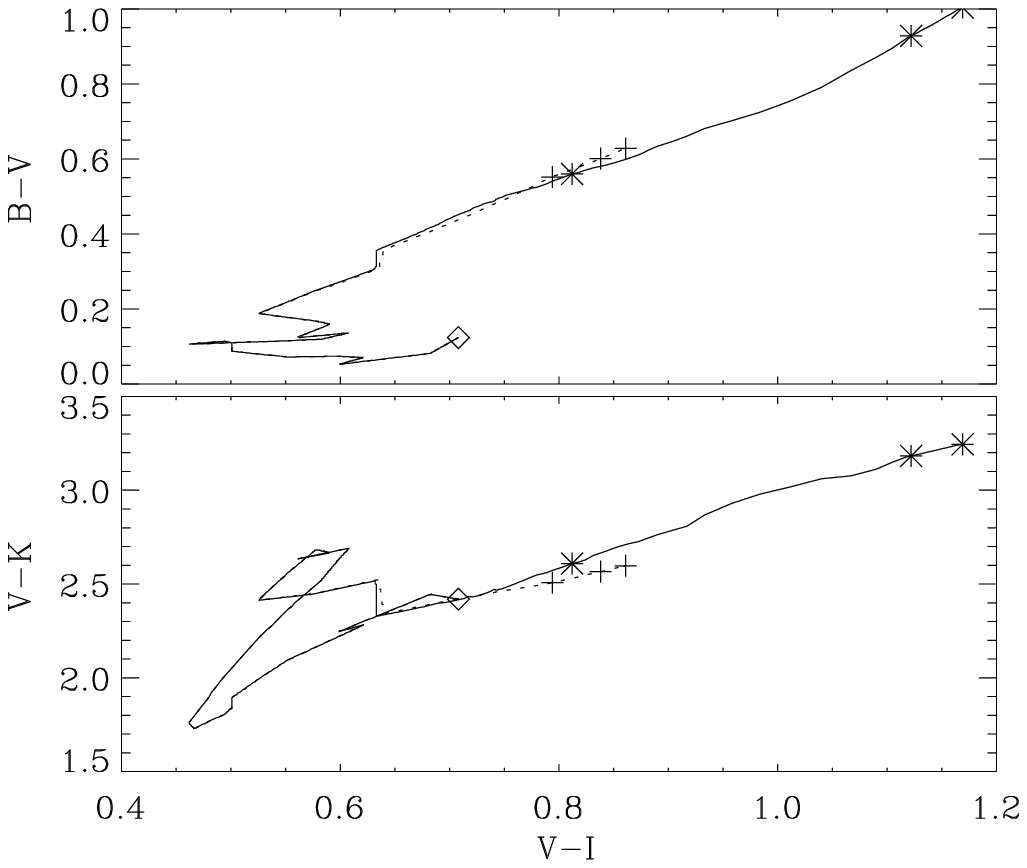,width=12.5cm}}
\vspace{-2.5cm}
\centerline{\hspace{+4.0cm}\psfig{figure=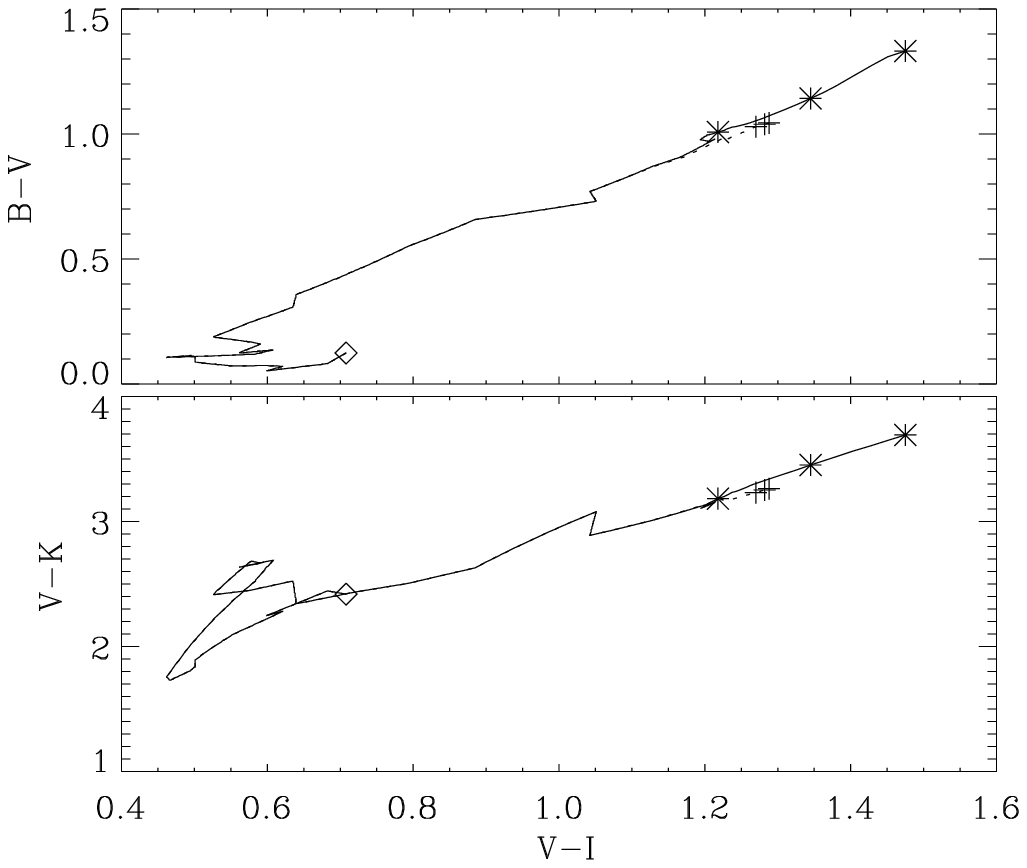,width=12.5cm}}
\vspace{-3.0cm}
\caption[]{The evolution of dissolving star clusters in the
            colour-colour diagram of the {\sl HST}/WFPC2 broad-band filters
$B-V$ vs $V-I$ and $V-K$ vs $V-I$, for clusters without initial
mass segregation ($\fsegr=0.0$ and $\tsegr=0.2$). Upper panel:
$\tdis=1$ Gyr; lower panel: $\tdis=10$ Gyr. The solid line shows the
colour-colour evolution with mass segregation and the preferential loss
of low-mass stars and the dashed lines show the evolution without the
loss
of low-mass stars, i.e. the evolution of the {\sc galev} models.
The diamond shows the colours in the beginning, at $t=12$ Myr. The
three asterisks show the colours of the cluster at three different ages:
$t / \tdis = 0.80$, 0.90 and 0.95 (reddest point). The three crosses 
show the colours of the standard {\sc galev} models at the same time. Notice that the
colours get much redder near the end of the lifetime of the clusters.}
\label{fig:model_colcol}
\end{figure}

The key question is; can we distinguish this effect on the basis of the
location of clusters in colour-colour diagrams?
Figure \ref{fig:model_colcol} shows the evolution of the dissolving
clusters in two colour-colour plots of $B-V$ vs $V-I$ and $V-K$ vs
$V-I$ for the same two models as shown in Fig.
\ref{fig:model_mass_phot}, i.e. with $\tdis=1$ and 10 Gyr, respectively.
The reddening of the clusters after $t>0.8~\tdis$ is highlighted
by marking the colours at $t=0.80$, 0.90 and 0.95~\tdis.
The difference is $\Delta(V-I)=0.3$ and 0.2 mag for the models with
 $\tdis=1$ and 10 Gyr, respectively.

Notice that in all colour-colour plots the reddening of the
clusters due to the loss of low-mass stars occurs almost along the lines of 
the normal colour-colour history (dashed lines in
Fig. \ref{fig:model_colcol}). 
{\it This implies that the effect of the preferential loss of low-mass
stars cannot easily be distinguished on the basis of colour-colour
plots. Hence, it could lead to errors in the age determination of 
unresolved clusters.}

\begin{figure}
\centerline{\psfig{figure=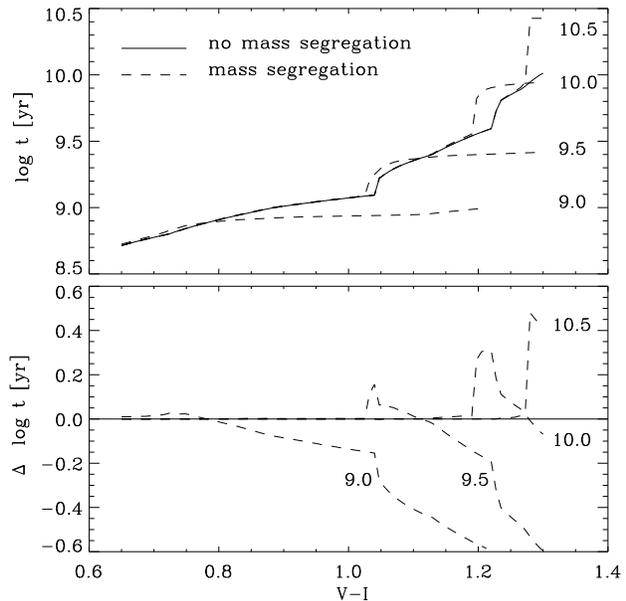,width=11.5cm}}
\caption[]{Upper panel: the relation between $V-I$ and the corresponding age of a
  cluster. The solid line is for clusters without mass segregation. The
  dashed lines are for clusters with mass segregation and the
  preferential loss of low-mass stars, for different dissolution times.
The parameters $\log (\tdis)$ are indicated in the panel. Lower panel: the logarithmic
  age correction to be made to ages derived from standard cluster
  evolution models.
}
\label{fig:age_colour}
\end{figure}

Since clusters with mass segregation evolve  almost along the same
lines as the standard cluster models in colour-colour diagrams, we
can plot the expected error in the age determination versus one of the colours.
The size of this error in the age estimate is 
shown in Fig. \ref{fig:age_colour}. In the upper panel we plot the age of a
cluster versus its $V-I$ colour. The solid line shows the relation
for the {\sc galev} cluster evolution models without mass segregation. The
dashed lines show the predicted colour-age relations for clusters
where mass segregation results in the preferential loss of low-mass 
stars. 
Notice that the age estimated on the basis of standard cluster models
without mass segregation will be roughly correct for the first 40\%
of the cluster's lifetime, i.e. for $\log(t) < \log(0.4 \tdis) = \log(\tdis) - 0.4$.
After that time a cluster with mass
segregation first gets slightly bluer than its counterpart without mass
segregation, so the age will be {\it underestimated} when conventional
cluster evolution models are used. This correction is about 0.15 dex in $\log(t)$
for clusters with $\tdis \simeq$ 3 Gyr, 0.30 dex for $\tdis \simeq$
10 Gyr and 0.5 dex for $\tdis \simeq$ 30 Gyr (lower panel
of fig \ref{fig:age_colour}).
This situation changes
during the last 20\% of the lifetime of the cluster when mass
segregation and the preferential loss of low-mass stars make the
cluster much redder than in conventional models. During that time, the cluster age will
be seriously {\it overestimated} when cluster models without mass
segregation are used to derive the age from the colours.
The age correction during that time reaches values as large as $-0.6$ dex.
 
We conclude that the ages of clusters derived from the broad-band
$B,~V,~I$ and $K$ colours using standard cluster evolution models
may be affected by significant systematic errors. During 
the first 40\% of the lifetime of a cluster its age will be 
about correct. During $0.4  < t /\tdis < 0.8$ the age will be 
{\it underestimated}.  During
the last 20\% of the cluster's lifetime the age will be overestimated
by a factor that rapidly increases towards the end of its lifetime.

(We have also calculated the photometric history of clusters with
primordial mass segregation, i.e. with $\tsegr=0$,
but these models are very similar to those without initial
mass segregation. If there is no primordial mass segregation, this will be
established within a short fraction, $\sim$ 20\%, of the cluster's
lifetime.)

\section{Discussion}

We have studied the effect of mass segregation and the preferential
loss of low-mass stars on the photometric evolution of unresolved star
clusters. The following assumptions were made:\\
(a) We adopted the {\sc galev} cluster evolution models, which are based on the Padova
evolutionary tracks that include a desription of the AGB evolution.
This is important because stars in the AGB phase
are the main contributors to the colour and brightness in 
long-wavelength bands after about 0.1 Gyr. 
We adopted the models for solar metallicity and
for a Salpeter IMF with index $-2.35$ down to an inital lower-mass limit
of $M_{\rm min}=0.15$ \Msun.
Mass loss due to stellar evolution is fully taken into account in our models.\\
(b) We simulated the effect of the preferential loss of low-mass stars
due to mass segregation in a simple way, by gradually removing the
stars with the lowest-mass stars. 
This is a simplification of the 
expected changes  in the stellar IMF due to the preferential loss of
low-mass stars (see Fig. 7 of BM2003).   This is not a bad
approximation, because we have tuned the increasing low-mass limit to 
the results of the $N$-body simulations in such a way that 
our model has the correct loss of low-mass stars due to evaporation. We have shown
that it results in a
predicted evolution of the {\it mean} stellar mass that is
very similar to that of the $N$-body simulations of BM2003.
 Since the mean stellar
mass after about 0.1 Gyr (i.e after the early high-mass loss phase of massive stars)
is mainly dependent on the cluster content of low-mass stars, the
agreement between our simpified model and the $N$-body simulations supports
the accuracy of our predictions.\\
(c) This first study is only for clusters with solar
metallicity. The same predicted effects can be expected
qualitatively for lower-metallicity clusters. In a follow-up study we
will extent the calculations to a large range of metallicities (Lamers \& Anders, in prep.).\\
(d) $N$-body simulations of BM2003 without initial mass segregation
show that cluster-wide mass segregation occurs after about 20\% of
the total lifetime of a cluster. On the other hand, observations of
very young clusters in the solar neighbourhood (Schilbach et al. 2005)
and in the Magellanic Clouds (de Grijs et al. 2002c and Sect. 2) show
that a fraction 
of the clusters are born with initial mass segregation, i.e.
with the massive stars concentrated towards the center and the lower-mass 
stars in the periphery. Therefore, we have done two sets of
calculations, one with initial mass segregation and one with mass
segregation after 20\% of the total lifetime of the cluster. \\
 (e) In our descriptive model, the cluster loses stars of all
masses proportional to its mass function in the initial phase at $t<\tsegr$
(i.e. before mass segregation has occurred).
This means that the slope of
the IMF does not change during this phase. 
After mass segregation has been established only the lowest-mass stars are
ejected from the cluster in our model. This results in a gradual shift
of the lower stellar mass limit. During all phases stellar evolution
removes the most massive stars. These two effects result in a narrowing
of the mass function. In our model the upper mass limit and the lower
mass limit meet each other at the moment the cluster is completely
dissolved. This is close, but not identical, to the predicted changes in
the mass function in the $N$-body simulations by BM2003.\\
 (f) We assume that stars which are lost from the cluster do not contribute any
longer to the photometry of unresolved clusters. 
The validity of this
assumption depends on the spatial resolution of the photometry,
i.e. on the point spread function (PSF) of the observations of unresolved
clusters, and on the speed with which stars leave the cluster.
For a simple estimate we assume that the stars leave the cluster typically
with a velocity on the order of the escape velocity. In that case it is easy to
show that the time it takes for a star to move beyond the radius of the PSF is 
$1 \times 10^7 r_{\rm PSF} \sqrt{R_{\rm cl}/M_{\rm cl}}$ yr, when $r_{\rm PSF}$
and $R_{\rm cl}$ are the radius of the PSF and the cluster,
  respectively, in pc and 
$M_{\rm cl}$ is the cluster mass in \Msun, if the path of the star is
perpendicular to the line of sight. For average directions this time
will be about $4/\pi$ as long. For a typical cluster of $10^4$ \Msun\
and $R_{\rm cl}=3$ pc an ejected star will leave the PSF with a radius
of 1 pc  (corresponding to $0.04''$  at $d=5$ Mpc) 
in about $2 \times 10^5$ yr. Therefore, unless the PSF covers a large area of
many pc$^2$, our assumption is reasonable.\\

With these assumptions  we predicted the photometric evolution of
unresolved star clusters and found that in the age range of $0.4 
\lesssim t/\tdis \lesssim 0.8$ the cluster will be bluer than
predicted by standard cluster evolution models. In the age range of $t \greatsim 0.8
\tdis$
the clusters will be redder than predicted by standard models.

The discovery of a group of apparent intermediate-age 
($2 - 5$ Gyr old) globular clusters in the
giant elliptical galaxy NGC 4365 by Forbes (1996), Gebhardt \&
Kissler-Patig (1999) and Larsen et al. (2001) may be the result of
mass segregation and the preferential loss of low-mass stars.
The luminosity-weighted age of clusters in NGC 4365 is about 14 Gyr
and there is no evidence of a recent merger. Another puzzling effect
is the lack of field stars in the age range from $2 - 5$ Gyr. 
The clusters were
reanalysed by Brodie et al. (2005) and Larsen et al. (2005). They
confirm the intermediate age based on the broad-band colours. However,
when Lick-indices are used to derive the ages of these clusters, the 
ages turn out to be between  10 and 14 Gyr, as expected for this
galaxy. Interestingly, the apparent intermediate-age clusters are
concentrated more towards the center of the galaxy than the other
clusters. 

We suggest that the apparent intermediate age clusters in NGC 4365 may,
in fact, be old clusters, $\sim$ 13 Gyr,  that have lost a significant fraction of their
low-mass stars due to mass segregation and tidal stripping. If this is
the correct explanation, then the age of the clusters is now between
about 40 and 80\% of their total lifetime, which implies that their
total lifetime is expected to be about 16 to 32 Gyr. 
These are reasonable values for the dissolution times of massive
clusters (BM2003). The concentration
of these clusters towards the center of the galaxy might be explained
by the fact that the total cluster lifetime, \tdis, is shorter close
to the centers of galaxies than far from the center. In fact, BM2003
have shown that for galaxies with a logarithmic gravitational potential
the dissolution time scales linearly with the distance to the galactic center
(except for interacting galaxies). The clusters in NGC 4365 that do
not show this apparent intermediate age are, on average, at greater
distances from the galactic center, so they have longer total
lifetimes. Their present age of 13 Gyr is too small to exhibit the
effects of mass segregation already.

In this explanation we did not discuss the effect of metallicity,
which also plays a role, both in stellar mass loss and in the colour
evolution, 
because the present study was done for clusters with solar metallicity only.
In our follow-up study (Lamers \& Anders, in prep.) we will predict 
the effects of mass segregation and
the preferential loss of low-mass stars for a large range of
metallicities. 

\section{Conclusions}

We predicted the photometric evolution of unresolved 
clusters with different total life times, \tdis, in the $B,~V,~R,~I$ and $K$ bands
of the {\sl HST}/WFPC2 broad-band filter system, with mass loss due to
stellar evolution and evaporation of low mass stars taken into account.
The photometric evolution
is compared with that of cluster models not including mass segregation nor
the preferential loss of low-mass stars. We call these the
``standard'' models. We considered clusters with a total lifetime in
the range of 0.3 to 30 Gyr. We obtained the following results.\\

\begin{enumerate}
\item{} During the first part of the lifetime of the cluster,
 i.e. during the
  first $\sim$ 40\%, irrespective of the total life time \tdis, the photometric
  evolution is the same as predicted for the standard models if the
  decreasing mass is taken into account. The dissolution of the
 cluster makes it fainter in all bands, but the colours are unaffected.
\item{} Between $\sim$ 40 and $\sim$ 80\% of its total lifetime the cluster is
{\it bluer} than predicted by the standard models. This is due to the
fact that the cluster has lost a large fraction of its (red) low-mass stars. 
This effect is
small: $\Delta (V-I)\simeq 0.03$ mag for clusters with $\tdis=1$ Gyr,
but increases steeply with increasing \tdis. It is about 0.1 mag for $\tdis=10$
Gyr. This implies that the age of the
clusters will be {\it underestimated} when standard models are used.
The error in the age estimate is about 0.15 dex in $\log(t)$ if $\tdis=3$ Gyr, 0.30
dex if $\tdis=10$ Gyr and 0.5 dex if $\tdis=30$ Gyr. Thus, the age of clusters with a
total lifetime of 20 Gyr and a real age of 14 Gyr, will erroneously be estimated 
as about 4 Gyr on the basis of the $V-I$ and $V-K$ photometry.
\item{} Between about $\sim$ 0.80 and 1.0 \tdis\ the clusters are much
  {\it redder} than predicted by the standard models. At those ages
  the AGB stars are the dominant contributors to the photometry at
  long wavelengths. These AGB stars are redder than the stars at the low-mass 
  end of the main sequence. So the removal of the lowest mass
  main-sequence stars will make the cluster redder than predicted by
  standard models. This effect increases from $\Delta(V-I)\simeq 0.0$ mag 
 at $t \simeq 0.8 \tdis$ to 0.3 mag at 0.95 \tdis. This reddening will result
  in an {\it overestimate} of the age of clusters based on broad-band
  photometry from $B-V$ to $V-K$ colours if  standard cluster evolution
  models are used. The effect is large and can grow to an overestimate
  of a factor $\sim$ 4 near the end of the cluster's life.
\item{} The changes in colour due to mass segregation and the
  preferential loss of low-mass stars occurs almost along the same
  lines in the colour-colour plots as the photometric evolution of the
  standard models. This makes it difficult to distinguish this effect
  from reddening due to the age of clusters.
\item{} The predicted photometric history of clusters with inital 
cluster-wide mass segregation is indistinguishable from that of
clusters without initial mass segregation. This is due to the fact
that mass segregation will occur quite rapidly
(within $\sim$0.20 \tdis), even if there were no initial
mass segregation. Because both the total lifetime of a cluster and the time for
mass segregation depend on the half-mass relaxation time, mass
segregation will occur at about a constant fraction of \tdis.  
\end{enumerate}

$N$-body simulations have shown that mass segregation and the
preferential loss of low-mass stars will occur in clusters in tidal
fields. Even if there is no initial mass segregation, the effects
predicted by our models will occur in real clusters.


\section*{Acknowledgements}

We thank the International Space Science Institute (ISSI) in Bern,
Switzerland, for hosting three star cluster workshops, in 2004 and
2005, when this project was carried out. 
We thank Mark Gieles, Marcel Haas, Diederik Kruijssen and Simon
Portegies Zwart for comments and suggestions.



\begin{thebibliography}{}

\bibitem{} 
Anders, P., Fritze-v. Alvensleben, U., 2003, A\&A 401, 1063 ({\sc
  galev} models)
\bibitem{}
Baumgardt, H., Makino, J., 2003, MNRAS, 340, 227 (BM2003)
\bibitem{} Bonnell, I.A., Davies, M.B., 1998, MNRAS, 295, 691
\bibitem{} 
Boutloukos, S.G., Lamers, H.J.G.L.M., 2003, MNRAS, 338, 717 
\bibitem{} 
Brandl, B., Sams, B.J., Bertoldi, F., Eckart, A., Genzel, R.,
Drapatz, S., Hofmann, R., L\"owe, M., Quirrenbach, A., 1996, ApJ, 466, 254
\bibitem{} 
Brandl, B., Brandner, W., Eisenhauer, F., Moffat, A.F.J.,
Palla, F., Zinnecker, H., 2001, in: Extragalactic Star Clusters, IAU
Symp. 207, eds. Grebel, E.K., Geisler, D., (ASP: San Francisco),
p. 226
\bibitem{}
Brodie, J.P., Strader, J., Denicoló, G., Beasley, M.A., Cenarro, A.J.,
Larsen, S.S., Kuntschner, H., Forbes, D.A., 2005, AJ, 129, 2643
\bibitem[1993]{bruzual}
Bruzual, G. A., Charlot, S., 1993, ApJ, 405, 538
\bibitem{} 
Campbell, B., et al., 1992, AJ, 104, 1721
\bibitem{} de Grijs, R., Johnson, R.A., Gilmore, G.F., Frayn, C.M.,
2002a, MNRAS, 331, 228
\bibitem{} de Grijs, R., Gilmore, G.F., Johnson, R.A., Mackey, A.D.,
2002b, MNRAS, 331, 245
\bibitem{} de Grijs, R., Gilmore, G.F., Mackey, A.D., Wilkinson, M.I.,
Beaulieu, S.F., Johnson, R.A., Santiago B.X., 2002c, MNRAS, 337, 597
\bibitem{} Elson, R.A.W., Tanvir, N., Gilmore, G.F., Johnson, R.A.,
Beaulieu, S.F., 1999, in: New Views of the Magellanic Clouds, IAU
Symp. 190, Chu, Y.-H., Suntzeff, N., Hesser, J., Bohlender, D., eds.,
Victoria, Canada, p. 417
\bibitem{} Fischer, P., Pryor, C., Murray, S., Mateo, M., Richtler, T.,
1998, AJ, 115, 592
\bibitem{}
Forbes, D.A., 1996, AJ, 112, 954
\bibitem{}
Gebhardt, K., Kissler-Patig, M., 1999, AJ, 118, 1526
\bibitem{} 
Gnedin O.Y., Ostriker J.P., 1997, ApJ, 474, 223
\bibitem{} 
Grebel, E.K., 2004, in: The Formation and Evolution of
Massive Young Star Clusters, ASP Conf. Ser., 322, Lamers, H.J.G.L.M.,
Smith, L.J., Nota, A., eds., (ASP: San Francisco), p. 101
\bibitem{} 
Gouliermis, D., Keller, S.C., Kontizas, M., Kontizas, E.,
Bellas-Velidis, I., 2004, A\&A, 416, 137
\bibitem{} Hillenbrand, L.A., 1997, AJ, 113, 1733
\bibitem{} Hillenbrand, L.A., Carpenter, J.M., 2000, ApJ, 540, 236
\bibitem{} Hillenbrand, L.A., Hartmann, L.E., 1998, ApJ, 492, 540
\bibitem{}
Hoogerwerf, R., de Bruijne, J.H.J., de Zeeuw, P.T., 2000, ApJ,
544, L133
\bibitem{} 
Hunter, D.A., Shaya, E.J., Holtzman, J.A., Light, R.M., O'Neil,
E.J., Lynds, R., 1995, ApJ, 448, 179
\bibitem{}
Hurley, J.R., Pols, O.R., Tout, C.A., 2000, MNRAS, 315, 543
\bibitem{} Inagaki, S., Saslaw, W.C., 1985, ApJ, 292, 339
\bibitem{} Kontizas, M., Hatzidimitriou, D., Bellas-Velidis, I.,
Gouliermis, D., Kontizas, E., Cannon, R.D., 1998, A\&A, 336, 503
\bibitem{} Kroupa, P., 2001, MNRAS, 322, 231 
\bibitem{}
Lamers, H.J.G.L.M., Gieles, M., Portegies Zwart, S.F., 2005a, A\&A, 429, 173
\bibitem{} 
Lamers, H.J.G.L.M., Gieles, M., Bastian, N., Baumgardt, H.,
Kharchenko, N.V., Portegies Zwart, S.F., 2005b, A\&A, 441, 117
\bibitem{}
Larsen, S.S., Brodie, J.P., Huchra, J.P., Forbes, D.A., Grillmaier,
C., 2001, AJ, 121, 2974
\bibitem{}
Larsen, S.S., Brodie, J.P., Starder, J., 2005, A\&A, 443, 413
\bibitem[1999]{leitherer} Leitherer, C., Schaerer, D., Goldader, J.D., et al., 1999, ApJS, 123, 3
\bibitem{} Larson, R.B., 1993, in: The Globular Cluster--Galaxy
Connection, Smith, G.H., Brodie, J.P., eds., ASP Conf. Ser. 48, (ASP:
San Francisco), p. 675
\bibitem{} Malumuth, E.M., Heap, S.R., 1994, AJ, 107, 1054
\bibitem{} N\"urnberger, D.E.A., Petr-Gotzens, M.G., 2002, A\&A, 382,
537
\bibitem{} Portegies Zwart, S.F., Makino, J., McMillan, S.L.W., Hut, P.,
1999, A\&A, 348, 117
\bibitem{} Salpeter, E., 1995, ApJ, 121, 161
\bibitem{} Santiago, B.X., Beaulieu, S., Johnson, R., Gilmore, G.F., 2001,
A\&A, 369, 74
\bibitem{} Schilbach, E., Kharchenko, N.V., Puskinov, A.E., R\"oser,
  S.  \& Scholz, R.-D, 2006, A\&A (in press) 
\bibitem{} Schulz, J., Fritze-v. Alvensleben, U., M\"{o}ller, C.S.,
Fricke, K.J., 2002, A\&A, 392, 1 ({\sc galev} models)
\bibitem{} Sirianni, M., Nota, A., De Marchi, G., Leitherer, C.,
Clampin, M., 2002, ApJ, 579, 275
\bibitem{} Stolte, A., Brandner, W., Brandl, B., Zinnecker, H.,
Grebel, E.K., 2004, AJ, 128, 765
\bibitem{} Sung, H., Bessell, M.S., 2004, AJ, 127, 1014 
\bibitem{} Testi, L., Palla, F., Prusti, T., Natta, A., Maltagliati, S.,
1997, A\&A, 320, 159
\bibitem{} Westerlund, B.E., 1961, Uppsala Astr. Obs. Ann., 5(1)
\end{thebibliography}
\end{document}